\documentstyle[epsf,12pt,world_sci]{article}   

\begin{document}

\title{\bf \large THE CONTEXT OF HIGH ENERGY QCD SPIN PHYSICS
\footnotemark[1]}

\author{\vspace{.2in}R. L. Jaffe\\[5ex]
{\small\em Center for Theoretical Physics} \\
{\small\em  Massachusetts Institute of Technology } \\
{\small\em                        Cambridge, MA ~02139~~U.S.A.}\\[1ex]
{\small\em and}\\[1ex]
{\small\em Department of Physics} \\
{\small\em  Harvard University} \\
{\small\em                        Cambridge, MA ~02138~~U.S.A.}}

\footnotetext[1]                                                               
{\baselineskip=16pt
This work is supported in part
by funds provided by
the U.S.~Department of Energy (D.O.E.)
under contracts \#DF-FC02-94ER40818 and \#DF-FG02-92ER40702
and by the National Science Foundation (N.S.F.) under
grant \#PHY 92-18167.
\hfil\break
MIT-CTP-2518 and HUTP-96/A010\hfil
January 1996\break
}

\maketitle

\vskip10ex

\centerline{Talk presented at the Adriatico Research Conference}
\centerline{Trends in Collider Spin Physics}
\centerline{ICTP, Trieste, Italy}
\centerline{5 -- 7 December 1995}
\centerline{and at the RIKEN Symposium}
\centerline{Spin Structure of the Nucleon}
\centerline{RIKEN, Wako-shi, Saitama, Japan}
\centerline{18 -- 20 December 1995}

\thispagestyle{empty}
\setcounter{page}{0}

\newpage
\bibliographystyle{unsrt}

As the first speaker I would like to give an broad view of the opportunities 
to explore QCD at a very high energy polarized hadron collider.  The better we 
understand perturbative QCD the more we are able to use it to probe the still 
unsolved mysteries of confinement.  My central message is that 
polarized colliders will provide a wealth 
of new information about the behavior of quarks and gluons inside hadrons 
complementing that available from lepton scattering and other more familiar 
probes of hadron structure.

My talk is organized as follows:
\begin{enumerate}
	\item Introduction
	\item Brief Remarks on the $g_1$ Situation
	\item Summaries of Some Issues Related 
		to Polarized Hadron Collider Physics
	\begin{itemize}
		\item Low Energy Flavor Physics
		\item Transverse Spin
		\item Higher Twist
	\end{itemize}
	\item Future Prospects at Polarized Hadron Colliders
	\begin{itemize}
		\item ${\cal A}_{TT}\over {\cal A}_{LL}$ --- a test of 
		the QCD Parton Formalism
		\item A Brief Summary of Some Flagship Experiments
	\end{itemize}
\end{enumerate}
\section{Introduction --- Precise Probes of Hadron Structure}
Light quark QCD looks simple and
elegant --- an unbroken non-Abelian gauge theory of nearly massless
($u$, $d$ and $s$) quarks interacting by 
with a Lagrangian one can fit on a postage
stamp, let alone a T-shirt,
\begin{equation}
	{\cal L}_{QCD}=-{1\over 4}\,{\rm Tr}\,{\bf F}_{\mu\nu}^2 +\bar q
	(i\gamma_\mu {\bf D}^\mu +m)q.
\end{equation}
Hadron phenomena are rich and regular, displaying many features
not accounted for by the symmetries of ${\cal L}_{QCD}$ alone. 
Examples include the OZI rule, the $SO(3)\times SU(6)$
classification of baryons, the absence of
exotics, and the early onset of asymptotic behavior in form factors and 
structure functions.  We have little fundamental 
understanding of these apparently simple 
regularities in terms of the underlying Lagrangian.  These regularities 
fascinate some of us even more than the computation of specific quantities 
like masses and coupling constants.

Traditional tools for exploring the structure of bound states have
only limited usefulness in the relativistic regime of light-quark
QCD.  Physicists unraveled atomic and nuclear
structure by precision studies of
i) excitation spectra, ii) electroweak form factors, and iii)
quasi-elastic scattering.  All three have yielded important
information about the structure of hadrons in the past --- the study
of hadron resonances gave us the quark model, form factor measurements
gave the first evidence of hadron compositeness, and quasi-elastic
lepton scattering of electrons and neutrinos led to QCD and the
quark-parton model.   However i) and ii) are considerably less
powerful in the relativistic regime than in traditional Schroedinger
quantum mechanics.  

From form factors we learn the single particle matrix elements of
electroweak currents.  ``Charges'' measured at zero momentum transfer
include the vector and axial charges of $\beta$-decay.
Just away from $\vec q = 0$ one obtains magnetic moments and charge radii,
$\langle r^2\rangle=-6{\partial F\over\partial q^2}\vert_{q^2=0}$,
{\it etc.\/}  The form factor of a non-relativistic system is directly
related to the fourier transform of some density in coordinate space.
However this connection fails beyond the first term or two
in the expansion in $q^2$, when the Compton wavelength
of a system is not negligible compared to its intrinsic
size.  For atoms and nuclei, $\lambda_c\ll \sqrt{\langle
r^2\rangle}$, but for the nucleon $\lambda_c/\sqrt{\langle
r^2\rangle}\approx 0.25$ and the connection is lost.  Despite this limitation, 
several interesting ``charges'' remain to be measured.
The programs to
measure the strangeness radius, axial charge and magnetic moment of
the nucleon at CEBAF show the vitality of this tradition, and connect the low 
energy physics community to the community interested in polarized collider 
physics.[\citenum{DonMus}] 

The interpretation of excitation spectra in QCD is complicated by the
fact that quarks and gluons are so light.  Even the first excited
state of the nucleon, the $\Delta(1232)$, is above particle (pion)
emission threshold.  The fundamental quanta --- quark and gluons ---
are so light that it is hard to imagine that hadrons are
well approximated as states of definite particle number.  Instead
modern quark models adopt a ``quasiparticle'' point of view, where
``constituent'' quarks and gluons are viewed as coherent states of the
fundamental quanta which preserve their identity in the hadronic
environment.  In general, spectra become uninterpretable in any
channel beyond the first couple of states.  Still there are several
fairly precise and very interesting questions in spectroscopy:  Where
are the ``glueballs''?  Where are the $CP$-exotic mesons that cannot be
non-relativistic $Q\bar Q$--states?  Is there a bound
or nearly bound dihyperon?

Quasielastic (deep inelastic) scattering from quarks and gluons, and its
generalizations to annihilation and hard hadron-hadron collisions
are much more important in QCD than in
nuclear or atomic physics.  In part this is because other tools are
less useful, but it also reflects the extraordinary power, precision
and flexibility the method attains in QCD.  Unlike the nuclear force,
QCD simplifies at short distances and the quasielastic approximation
({\it i.e.\/} the renormalization group improved parton model) becomes
exact at large momentum transfer.  Furthermore, corrections to
the leading asymptotic behavior (so-called ``higher twist'' effects)
can be classified and analyzed with the same tools.  Parton
distribution and fragmentation functions appear to be the most useful
and abundant information available on the structure of hadrons. 
Extensions to spin dependent and higher twist effects offer a wealth
of new information which can be the testing ground both of
phenomenological models of confinement and of numerical simulations.  

At present we have accurate information on the momentum
distribution of quarks and gluons in the nucleon.  We have fairly
accurate measurements of the helicity distribution of quarks in the
neutron and proton.  We know that at
$Q^2=20$ GeV$^2$ gluons carry about 43\% of the nucleon's momentum, quarks
carry about 57\%,[\citenum{MRS}]
$\bar u$ and $\bar d$ quarks carry $13.5\pm0.3\%$, and
$s$ and $\bar s$ quarks carry $4.1\pm 0.4\%$.[\citenum{CCFR,NLO}]  We
know that $\bar u$ and
$\bar d$ quarks are distributed differently within the proton.  We
know that the fraction of the spin of the nucleon carried on the spin
of the quarks is small --- the most recent value is about $20\pm 10\%$
at $Q^2=10 GeV^2$.  These are examples of the kind of precise
information obtained by using perturbative QCD to probe confinement.

In the not too distant future as a result of experimental programs
described at this meeting, we can expect to have information on
\begin{itemize}
	\item the flavor dependence of quark and antiquark helicity
	distributions,
	\item the helicity weighted distribution of gluons in a polarized
	nucleon,
	\item information on polarized quark-gluon correlation functions from
	the study of the transverse spin distribution, $g_2$,
	\item estimates of the {\it transversity\/} weighted distribution of
	quarks in the nucleon, and
	\item measurements of spin dependent effects in quark and
	gluon fragmentation processes.
\end{itemize}
Experiments and facilities at SLAC, CERN, HERA, and BNL all will
contribute to this program.  Other experiments at lower energies
at CEBAF, Mainz and Bates will provide complementary information on
the spin and flavor structure of hadrons.

The primary objects of interest in short distance probes of nucleon
structure are quark and gluon distribution functions.  Table~\ref{tbl:t2}
 summarizes all the leading twist (scaling, modulo
logarithms) distributions of quarks and gluons in a spin-$1/2$
target.  Some, like the transversity distribution, $\delta q$, cannot
be measured in conventional deep inelastic scattering and are prime
candidates for experiments at polarized hadron colliders.  Others like
the polarized gluon distribution,
$\Delta G$, may be more accessible in polarized hadron colliders than
in deep inelastic lepton scattering.  The twist three (${\cal
O}(1/Q)$, mod logs) distributions listed in Table~\ref{tbl:t3}
provide precise probes of quark--gluon correlations in the nucleon. 
While $g_T$ is most accessible in deep inelastic scattering, $h_L$
may be measured in polarized hadron collisions. 
\begin{table}[ht]
\center{
\begin{tabular}{|c|c|c|c|}
\hline
Parton & Spin Average & Helicity Difference  & Transversity Difference \\
& $f_1$ & $g_1$ & $h_1$ \\  \hline
Quark &  $q(x,Q^2)$ &$ \Delta q(x,Q^2)^\dagger$ & $ \delta
q(x,Q^2)^\dagger$\\ \hline Antiquark &  $\bar q(x,Q^2)$ &$ \Delta \bar
q(x,Q^2)^\dagger$ & $ \delta 
\bar q(x,Q^2)^\dagger$\\ \hline
Gluon &  $G(x,Q^2)$ & $ \Delta G(x,Q^2)^\dagger$ &  ---$\phantom{-}^\dagger$\\
\hline
\end{tabular}
} 
\caption{\sf Quark, antiquark and gluon distribution functions at leading 
twist.  Those marked with a $^\dagger$ are particularly interesting for
polarized hadron colliders.  Note the absence of a gluon transversity 
distribution at leading twist.}
\label{tbl:t2}
\end{table}
\begin{table}[ht]
\center{
\begin{tabular}{|c|c|c|c|}
\hline
Parton & Spin Average & Helicity Difference  & Transversity Difference
\\\hline 
Quark \& antiquark &  $e(x,Q^2)$ &$ h_L(x,Q^2)^\dagger$ & $
g_T(x,Q^2)$\\
\hline
\end{tabular} 
}
\caption{\sf Quark and antiquark distributions at 
twist-three ${\cal O}(1/Q)$.}
\label{tbl:t3}
\end{table}

\section{Brief Remarks on the $g_1$ Situation}
\begin{figure}[h]
\quad\begin{minipage}[t]{2.75 in}
\epsfxsize=2.75in
\epsffile[80 255 451 662]{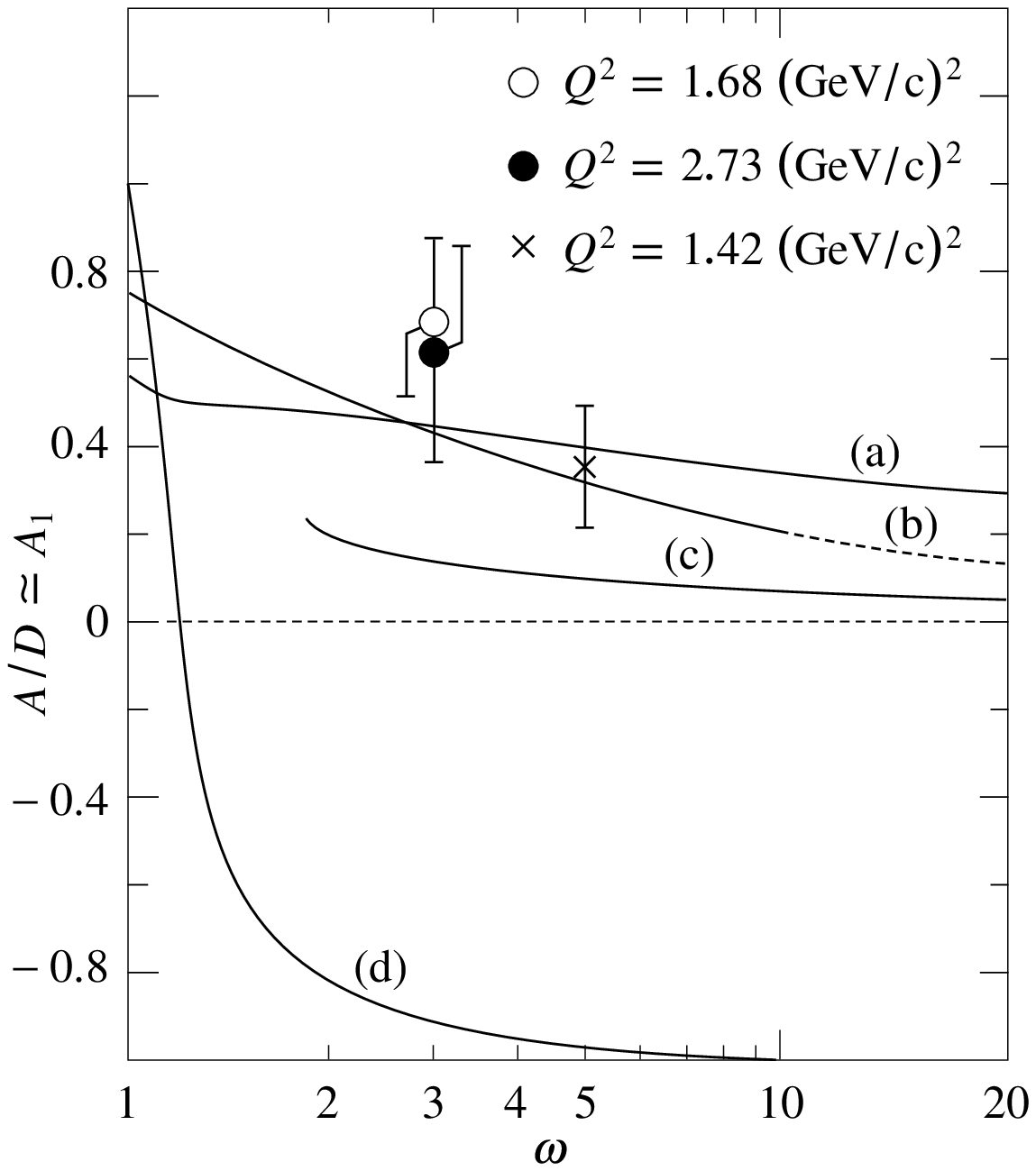}
\footnotesize \caption{\sf $A_1$ data from SLAC E-80 in 1976.}
\label{fig:E80}
\end{minipage}
\hfill
\begin{minipage}[t]{2.75 in}
\epsfxsize=2.5in\epsffile{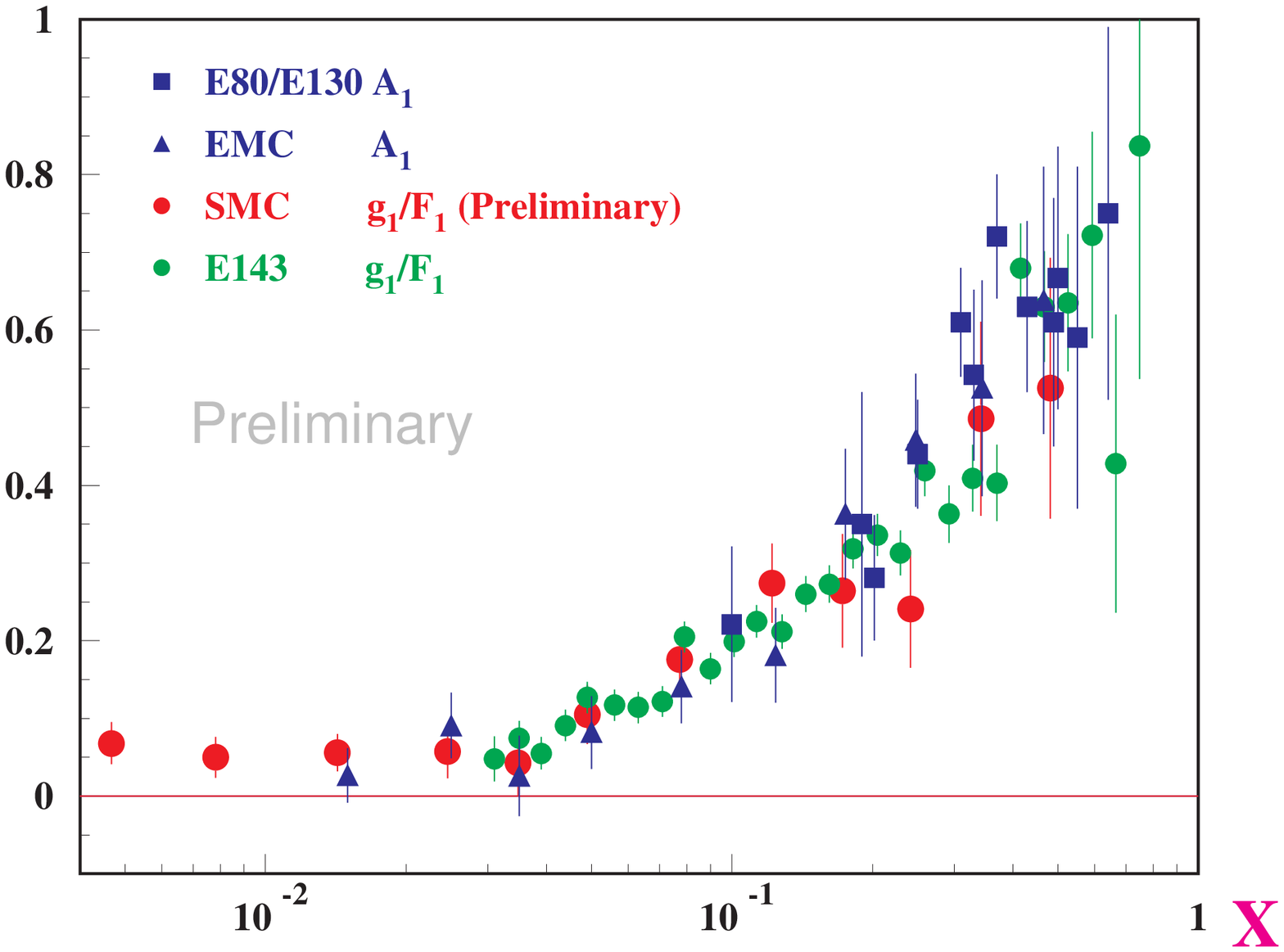}
\footnotesize \caption{\sf A recent compilation of data on $A_1$ from SMC.}
\label{fig:SMC-E143}
\end{minipage}\quad
\end{figure}

\begin{figure}[t]
\quad\begin{minipage}[t]{2.5 in}
\epsfxsize=2.5in
\epsffile{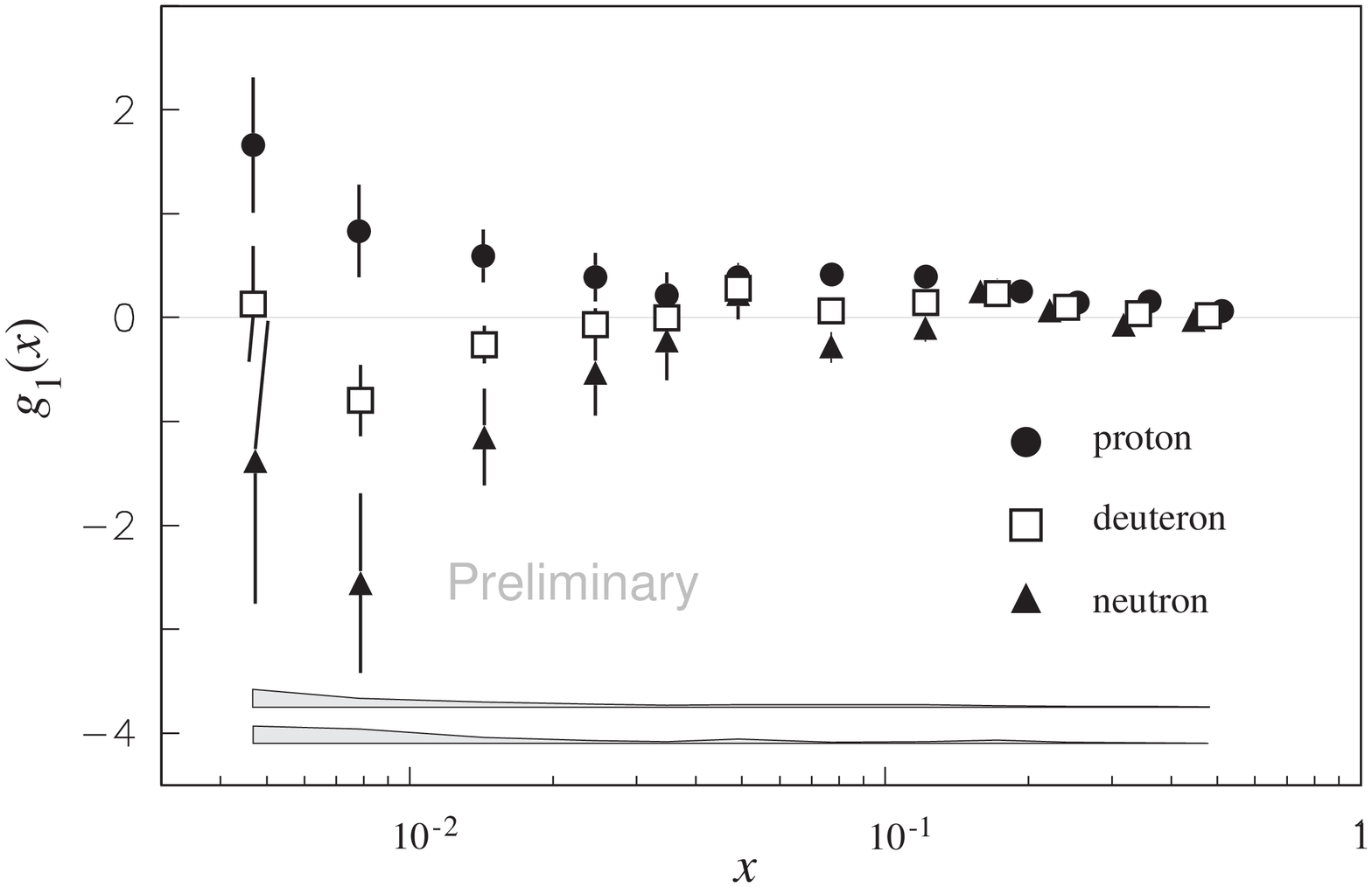}
\footnotesize \caption{\sf Data from SMC on $g_1$ for the proton, neutron and 
deuteron.}
\label{fig:g1pnd}
\end{minipage}
\hfill
\begin{minipage}[t]{2.5 in}
\epsfxsize=2.5in\epsffile[21 252 547 587]{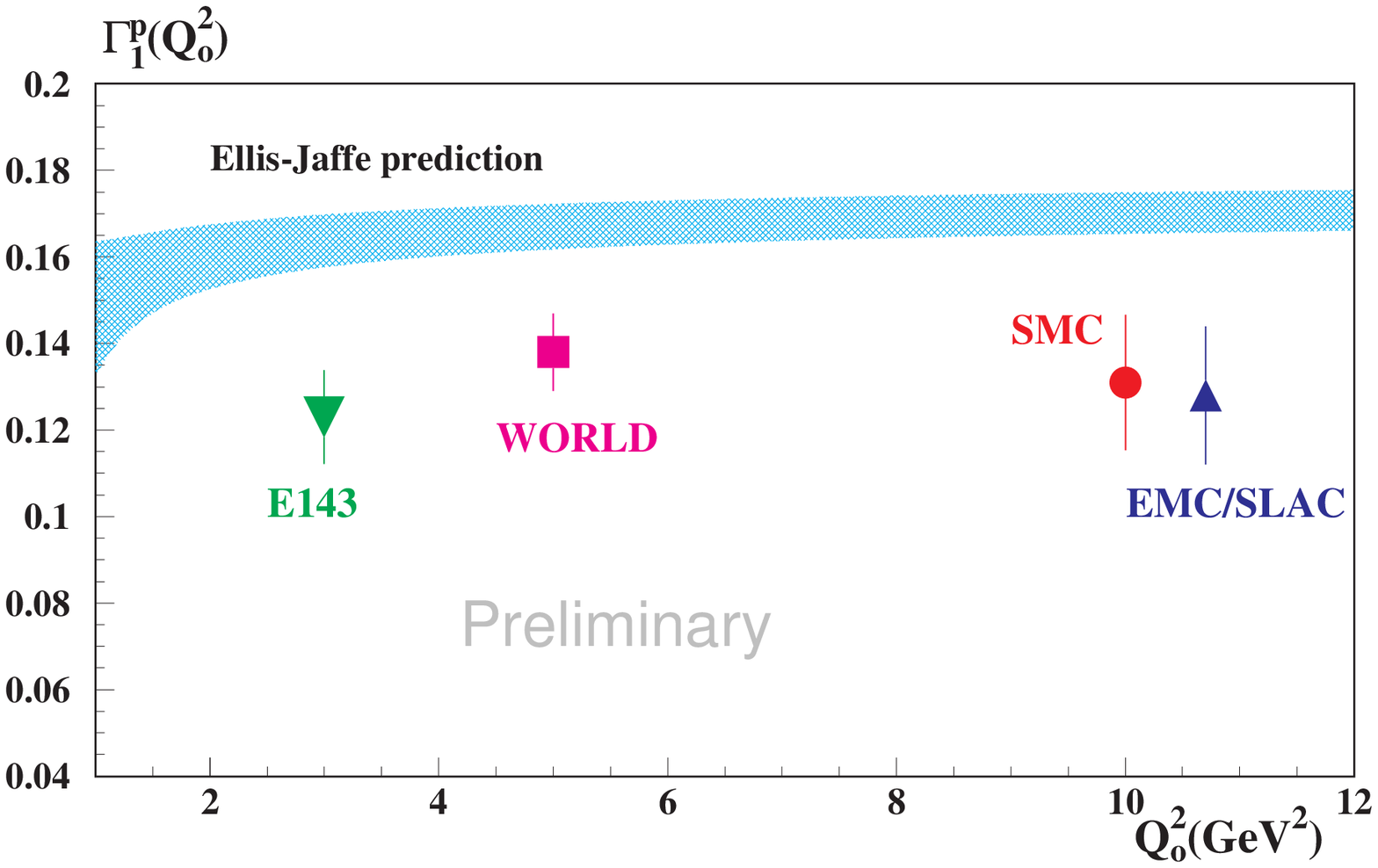}
\footnotesize \caption{\sf Comparison of data with the assumption $\Delta s=0$ 
from SMC.}
\label{fig:EJdata}
\end{minipage}\quad
\end{figure}

It is impossible to give a talk on the future prospects for probing
the spin structure of hadrons without mentioning the tremendous
progress that has been made since the first attempts to measure the
deep inelastic spin asymmetry at SLAC in the early 1970's.
Fig.~\ref{fig:E80} shows the data published in 1976 by the SLAC/Yale
(E80) collaboration.  For comparison Fig.~\ref{fig:SMC-E143} shows a
compilation of asymmetry data including recent CERN (SMC) and SLAC
(E143) measurements with the old SLAC/Yale (E80/130)
data.\footnote{Figs.~\ref{fig:SMC-E143}, \ref{fig:g1pnd} and
\ref{fig:EJdata} are copies of the transparencies shown by
J.~Lichtenstadt at the Trieste Conference.  They are preliminary and
subject to revision.  I am grateful to Dr.~Lichtenstadt for copies of his
figures.}  A recent compilation of data on $g_1$ for proton, neutron and
deuteron targets is shown in Fig.~\ref{fig:g1pnd}.  Such accurate data
has allowed experimenters to test the sum rules that relate integrals
over data to forward matrix elements of axial charges. These are
perhaps the simplest examples of the sort of precise information
available from deep inelastic probes.

The Bjorken Sum Rule follows from isospin symmetry and the short distance
analysis of QCD.[\citenum{Bj66}]  So firm a dynamical basis has both advantages
and disadvantages.  On the one hand, the BjSR tests QCD at a very fundamental
level.  On the other hand, if it is verified, we have learned nothing new
about hadron structure.  The elementary form of the sum rule reads
\begin{equation}
	\int_0^\infty dx g_1^{p-n}(x,Q^2) = {1\over 6}{g_A\over g_V},
	\label{eq:BJeasy}
\end{equation}
where ${g_A\over g_V}$ is the ratio of axial vector decay constants
measured in $n\rightarrow p e^-\nu$. 
Although I will generally ignore the real complexity that underlies
comparison between theory and experiment in QCD, the BjSR provides an
opportunity to illustrate the effects that must be taken into account if a
reliable comparison is to be made.  Three conceptually different types of
corrections afflict eq.~(\ref{eq:BJeasy}):  i) perturbative QCD corrections to
the leading twist operator  --- the isovector axial charge; ii) kinematic
${\cal O}({1\over Q^2})$ corrections due to the non-vanishing mass of the
target nucleon; and iii) dynamical ${\cal O}({1\over Q^2})$ due to
quark-gluon corrections in the nucleon --- the so-called higher-twist
effects.  These replace the original sum rule with,
\begin{eqnarray}
\int_0^\infty dx g_1^{p-n}(x,Q^2) &=& {1\over 6}{g_A\over g_V}
	\{1-{\alpha_s\over\pi}-{43\over 12} {\alpha_s\over\pi}^2 - 20.215
	{\alpha_s\over\pi}^3 + \ldots\}\nonumber \\
&+&	{M^2\over Q^2}\int_0^1 dx x^2\{{2\over 9}g_1^{p-n}(x,Q^2)+{1\over 6} g_2
	^{p-n}(x,Q^2)\}\nonumber\\
&-&	{4\over 27} {1\over Q^2}{\cal F}^{u-d},
	\label{eq:BJreal}
\end{eqnarray}
where ${\cal F}$ is the invariant matrix element associated with a
twist-four operator,
$	2{\cal F}^uS^\sigma = \langle P,S\vert g \bar u \tilde
	F^{\sigma\lambda}\gamma_\lambda u\vert P,S\rangle$.
These corrections complicate the comparison of experiment with theory.  On the 
other hand, if $g_1^{p-n}$ can be measured accurately enough, the
magnitude of this quark-gluon correlation within the nucleon can be
extracted from experiment.  

The separate spin sum rules for the proton and neutron are not as fundamental 
as the BjSR because they involve an otherwise unknown nucleon axial charge.  
If we define the light quark axial charges of the proton by
$\Delta q(Q^2)S_\alpha=\langle P,S\vert\bar 
	q\gamma_\alpha\gamma_5q\vert P,S\rangle$, for $q=u,d,s$, then we can write sum
rules for the proton and deuteron  (ignoring any bound state corrections for
the deuteron),
\begin{eqnarray}
	\int_0^1 dx g^{ep}(x,Q^2) & = &{1\over 2}\left({4\over 9}\Delta u(Q^2)
	+ {1\over 9}\Delta d(Q^2) + {1\over 9}\Delta s(Q^2)\right)\nonumber\\
	& = & {1\over 18}\left( 3F + D + 2\Sigma(Q^2)\right)
	= {1\over 18}\left( 9F - D + 6\Delta s(Q^2)\right)\nonumber\\
\int_0^1 dx g^{ed}(x,Q^2) & = &{1\over 2}\left({5\over 9}\Delta u(Q^2)
        + {5\over 9}\Delta d(Q^2) + {2\over 9}\Delta s(Q^2)\right)\nonumber\\
        & = & {1\over 18}\left( 3F - D + 4\Sigma(Q^2)\right)
	= {1\over 18}\left( 15F - 5D + 12\Delta s(Q^2)\right).\nonumber\\
\label{ejsr}
\end{eqnarray}
All perturbative QCD, target mass and higher twist corrections have been 
suppressed in eqs.~(\ref{ejsr}).  Two linear combinations of the three light 
flavor axial charges can be related to (scale independent) combinations 
of the axial charges ($F$ and $D$)
measured in hyperon and neutron $\beta$--decay.  The third, $\Sigma$, is the 
fraction of the spin of the nucleon carried by the spin of the light quarks, 
and depends on $Q^2$ because the conservation of the associated axial current 
is ruined by the axial anomaly.
\begin{eqnarray}
	\Delta u(Q^2) - \Delta d(Q^2) &=& F + D\nonumber\\
	\Delta u(Q^2) + \Delta d(Q^2) - 2\Delta s(Q^2) & = & 3 F -D\nonumber\\
	\Delta u(Q^2) + \Delta d(Q^2) + \Delta s(Q^2) & = & \Sigma(Q^2)
	\nonumber\\
\label{delta}
\end{eqnarray}
These relations assume exact $SU(3)_{\rm flavor}$ 
symmetry for octet baryon axial currents, 
which seems to agree with available data (although see below for further 
discussion).

When eqs.~(\ref{ejsr}) were first derived, $\Delta s$ was assumed to 
vanish.[\citenum{EJ}]  Although this may have been reasonable 
at the time, we now know that each of 
the light-flavor axial charges is 
scale dependent (via the anomaly [\citenum{Kod}]) 
so if $\Delta s$ vanishes at one $Q^2$ it will not vanish elsewhere.  It is 
interesting to note, nevertheless, that the assumption $\Delta s= 0$ 
corresponds to $\Sigma = 3F - D$, and with the current values of $F+D=1.2573\pm 
0.0028$ and $F/D=0.575\pm 0.016$ one gets $\Sigma = 0.579\pm 0.025$.  
So $SU(3)_{\rm 
flavor}$ symmetry and no polarized strange quarks in the nucleon already 
suggest something peculiar going on with the nucleon spin.[\citenum{Seh}]
This is associated 
with the failure of the non-relativistic quark model to account for axial 
charges --- {\it viz.\/} the famous $g_A/g_V = 5/3$ for neutron $\beta$--decay.

Fig.~\ref{fig:EJdata} shows 
a recent compilation of the world's data on the proton sum rule compared with
the prediction $\Delta s = 0$.  Clearly there is strong evidence that $\Delta s$
is negative  and therefore that $\Sigma$ is even smaller than anticipated in the
quark  model.  Recent values hover around $\Sigma = 0.2\pm 0.1$ or,
equivalently, 
$\Delta s = -0.12 \pm 0.04$ using the relation $\Sigma = 3 F - D + 3 \Delta 
s$.[\citenum{SMCdeut}]

There have been many reviews of the issues raised by the unexpectedly small 
fraction of the spin of the nucleon carried on the spin of the light quarks.  
For an elementary introduction see Ref.~[\citenum{JafPT}].  Here,
briefly, are a few comments from a theoretical perspective:

\begin{itemize}
	\item	Theorists have not come up with a ``gee whiz'' solution,
		{\it i.e.\/} there is no simple and elegant explanation that
		leaves conventional quark model phenomenology intact and
		explains the small value of $\Sigma$.
	\item	Certain ``trivial'' effects raised when the data first appeared
		still plague the interpretation of the data.
		In particular,
	\begin{itemize}
		\item	Violation of $SU(3)_{flavor}$ symmetry for octet
			axial charges could affect the evaluation of the
			sum rules.  There has 
			been a recent flurry of activity
			on this subject.[\citenum{LL,Man95,Sch,forte}]  Forte
			has pointed out that the extraction of $\Sigma$ from
			the data is relatively insensitive to uncertainties
			in the less well known combination $3F-D$.[\citenum
			{forte}]  However his analysis stays within the
			parameterization (eq.~(\ref{delta})) given by
			exact SU(3) and therefore does not bound the effects 
			of SU(3) violation.  Refs.~[\citenum{Man95,Sch}] both
			claim that SU(3) violation can be sufficient 
			to give $\Delta s = 0$. Direct measurement of 
			$\Delta s$ in $\nu p$ {\it elastic\/} scattering
			as proposed at Los Alamos would settle the issue.
			[\citenum{LSND}]
		\item	Unanticipated behavior of the functions $xg_1(x,Q^2)$
			as $x\rightarrow 0$ could affect the sum rules.	
			The latest low--$x$ data from SMC (see 
			Fig.~\ref{fig:g1pnd}) suggest tantalizing 
			deviations from expected behavior --- note, for	
			example the increase of $g_1^d$ at the lowest-$x$.  
			Unfortunately it is hard to obtain much guidance from
			data with such large error bars.

		\item	No signs of strong, sub-asymptotic $Q^2$ dependence
			have been seen at SLAC.  The possibility that the
			early evaluations of the sum rules were contaminated
			by sub-asymptotic corrections [\citenum{AIL,EK}]
			that would vanish at
			higher $Q^2$ and reveal a larger value of $\Sigma$
			seems less likely as more accurate low $Q^2$ data
			and very large $Q^2$ data become available.[\citenum{EK2}]
	\end{itemize}
	\item  	What does carry the spin of the nucleon?  A separation into
		four terms is often quoted,
		\begin{equation}
			{1\over 2} = {1\over 2}\Sigma + \Delta L_Q + \Gamma + 
			\Delta L_G,
		\label{Jsr}
		\end{equation}
		where $\Delta L_Q$ and $\Delta L_G$ are the quark and gluon
		orbital angular momenta, and $\Gamma$ is the gluon spin. 
		[\citenum{JafMan}]  It is not 
		obvious that the total angular momentum can be grouped 
		into the sum of four terms.  After all, the energy cannot
		be written this way because it contains interaction 
		dependent terms that cannot be attributed to quarks or gluons
		separately.  Recently Ji and collaborators have shown that such
		a separation is both gauge and renormalization scheme 
		dependent.[\citenum{JTH}]  There is a natural definition of
		$\Gamma$ as the integral over $\Delta G$,[\citenum{Jafglue}]
		and Ref. [\citenum{JTH}] argues for particular definitions
		of $\Delta L_Q$ and $\Delta L_G$ with interesting consequences
		(see below).  However a definition of 
		$\Delta L_Q$ and $\Delta L_G$
		that makes contact with experiment is still lacking.

		Many have speculated that $\Gamma$ is large and 
		positive in order to make up for the deficiency in
		$\Sigma$.[\citenum{gluon}]  {\it A priori\/} not even
		the sign of $\Gamma$ is known, and it may be 
		negative.[\citenum{Jafglue}]
		This question can be addressed by experiment 
		since $\Gamma$ can be 
		defined in terms of an integral over the polarized gluon
		distribution in the nucleon[\citenum{Jafglue}] and can be 
		measured.  

	\item	Ji, Tang and Hoodbhoy have studied the $Q^2$ 
		evolution of the four terms in eq.~(\ref{Jsr}).[\citenum{JTH}]
		In light-cone
		gauge they can use traditional parton model analysis to
		calculate the GLAP splitting functions necessary to construct
		evolution equations.  Two of their observations are 
		particularly interesting:  

	\begin{itemize}
		\item  	First, they find that the anomalous
			dimension matrix for the four operators
			has a zero 
			eigenvalue, with the result 
			that in the {\it very\/} large
			$Q^2$ limit where all 
			other eigenvectors of the anomalous
			dimension matrix have evolved 
			to zero, the ratio of quark and
			gluon contributions goes to a 
			definite limit, independent
			of hadronic target.  This is 
			exactly analogous to the
			result for the momentum sum rule, 
			and indeed, the numbers are the same:
		\begin{eqnarray}
			\lim_{\ln Q^2\rightarrow\infty}
			\Delta L_Q + {1\over 2}\Sigma & =
			& {1\over 2}
			{3N_f\over 16 + 3 N_f}\nonumber\\
			\lim_{\ln Q^2\rightarrow\infty}
			\Delta L_G + \Gamma & = & {1\over 2} 
			{16\over 16 + 3 N_f}\nonumber \\
		\label{asymptote}
		\end{eqnarray}
			The result can be motivated by the observation that the
			angular momentum density tensor 
			is linearly related to the
			stress tensor --- 
			$M^{\mu\nu\lambda}=x^\nu T^{\mu\lambda}
			-x^\lambda T^{\mu\nu}$ up to total derivatives.
			For reasons that have never been well understood, the
			momentum is partitioned according to this  asymptotic
			result even at relatively low 
			$Q^2$ --- ${\varepsilon_Q\over
			\varepsilon_G}\approx 1$.  If the same is true for
			the angular momentum then,
			$\Delta L_Q + {1\over 2}\Sigma \approx {1\over 4}$ and 
			$\Delta L_G + \Gamma \approx {1\over 4}$
   even at moderate $Q^2$.
			Continuing this speculation to 
			its logical conclusion, we
			can combine this with the 
			known result that $\Sigma\approx 
			0.3$ and eq.~(\ref{Jsr}), to obtain,
			${1 \over 2}\Sigma \approx 0.15,
			\Delta L_Q \approx 0.1$ and
			$	\Gamma + \Delta L_G \approx 0.25$,	
			leaving only the 
			separation between $\Gamma$ and $\Delta L_G$ open
			to debate.  It would be very 
			interesting to find a way to test this speculation.

		\item	Second, they find that they can 
			study either of the
			two popular ways of regulating the infrared behavior
			of the quark spin contribution.[\citenum{Bod}]
			If they choose the
			prescription suggested by the operator product 
			expansion, say in $\overline{MS}$ scheme, then they
			obtain eq.~(\ref{Jsr}) itself.  If, however, they
			use an a scheme in which the gluonic contribution
			to the quark axial charges arising from the triangle
			anomaly is explicitly separated out (as, for example,
			advocated in Refs.~[\citenum{gluon}]), 	
			$\Sigma\rightarrow \Sigma-{\alpha\over 
				2\pi}N_f\Gamma$,
			then they find that the quark orbital angular momentum
			has a compensating gluonic contamination,
			$\Delta L_Q\rightarrow \Delta L_Q + {\alpha\over 4\pi} 
				N_f\Gamma$,
			so that eq.~(\ref{Jsr}) now reads,
			\begin{equation}
				{1\over 2} = {1\over 2}\left( 
				\Sigma -{\alpha\over 
				2\pi}N_f\Gamma\right) +\left( \Delta 
				L_Q + {\alpha\over
				4\pi}N_f\Gamma\right)+\Delta L_G +\Gamma.
				\label{Jsrnew}			
			\end{equation}
			So the net effect of the renormalization scheme 
			dependence is only to shift a contribution between
			quark spin and orbital angular momentum.  
	\end{itemize}
\end{itemize}

\section{Issues Related to Polarized Hadron Collider Physics}

Although the primary motivation for pursuing a program in polarized 
hadron-hadron physics at a very high energy collider comes from the surprises 
discovered in polarized deep inelastic scattering, there are several other 
areas of hadron physics that will contribute to or be affected by such a 
program.  As we discuss the physics opportunities at a polarized hadron 
collider at this workshop it is important to keep these connections in mind.

\subsection{Strange Quarks in the Nucleon}

The observation that strange quark matrix elements in the 
nucleon are larger than expected in the naive quark model began the revival of 
interest in quark and gluon substructure of hadrons.  The current state of 
affairs is summarized in Table~\ref{tbl:strange}.  
[For data and further discussion 
on $\varepsilon_s+\varepsilon_{\bar s}$
see Ref.~[\citenum{CCFR}], on $m_s\langle\bar s 
s\rangle$ see Ref.~[\citenum{Sainio}], on
$\Delta S$ see ref.~[\citenum{SMC}]]

\def\fixit#1{\lower1.5ex\hbox{#1}}
\begin{table}[ht]
\center{
\begin{tabular}{|c|c|c|c|c|}
\hline
Operator & Name & Quarks & Current Value & Comments\\  \hline
$s^\dagger s$ & {\footnotesize STRANGENESS} & $s-\bar s$ & 0 & 
Just an example \\
\hline
\lower1.5ex\hbox{$r^2s^\dagger s$} &
{\footnotesize STRANGE RADIUS} & 
\fixit{$s-\bar s$} & \fixit{???} & \fixit{$e^\uparrow p
\rightarrow ep$} \\
& $\langle r^2_s\rangle$ & & &\\
\hline
 & {\footnotesize STRANGENESS} & & 
 & $\nu N\rightarrow \mu^+ \mu^- X$\\
$\bar s \gamma^+\partial^+ s$ & {\footnotesize MOMENTUM FRACTION}
&$s+\bar s$ &$0.0408\pm 0.0041$&
$\bar\nu N\rightarrow \mu^+\mu^- X$\\ &
$\varepsilon_s+\varepsilon_{\bar s}$ &&& \\
\hline
 & {\footnotesize STRANGE} &  &
 & Low energy \\
$m_s\bar s s$&{\footnotesize SCALAR DENSITY}&$s+\bar s$ &$190\pm 60$
MeV&$\pi$N scattering\\ &$m_s\langle\bar s s\rangle$&&&\\
\hline
&{\footnotesize STRANGE} & 
$s^\uparrow-s^\downarrow$ & &\\
${1\over 2}\vec r\times s^\dagger\vec\alpha s$ & {\footnotesize MAGNETIC
MOMENT} & $+\bar s^\downarrow - \bar s^\uparrow$& ???& $e^\uparrow
p\rightarrow ep$\\ & $\mu_s$ &&&\\
\hline
 & {\footnotesize STRANGE} & $s^\uparrow
-s^\downarrow$ &  & $e^\uparrow p^\uparrow\rightarrow e X$,\\
$\bar s\vec\gamma\gamma_5 s$& {\footnotesize SPIN FRACTION} & $-\bar
s^\downarrow + \bar s^\uparrow$ &$-0.10\pm0.03 $ & 
$\nu N\rightarrow\nu N$\\
& $\Delta s$ &&& \\
\hline
& {\footnotesize STRANGE} & $s^\uparrow-\bar s^
\uparrow$ && Chiral odd deep \\
$i\bar s\sigma^{0i}\gamma_5 s$ &{\footnotesize TENSOR CHARGE} &
$-s^\downarrow+\bar s^\downarrow$ & ??? & inelastic processes\\
& $\delta s$ &&&\\
\hline
\end{tabular}
} 
\caption{\sf Current information on the 
nucleon matrix elements of $\bar s s$ 
operators. 
}
\label{tbl:strange}
\end{table}

Strange quark matrix elements in the 
nucleon give us the most precise measures of
OZI-Rule violation.  Naive 
quark models exclude strange quarks from the
nucleon.  Perturbative QCD, dynamical chiral symmetry breaking and
non-perturbative models of confinement 
all introduce $\bar s s$ admixtures.  Some
models, like the $SU(3)_{flavor}$ symmetric Skyrme 
model go to the opposite extreme from the naive quark model
and introduce very large strangeness admixtures into
 the nucleon.
I would like to stress a few points ---
\begin{itemize}
\item The strange quark contribution 
to the momentum fraction increased
significantly a couple of years ago
 with the realization that a (next
to leading order (NLO)) perturbative
 correction to $\nu p\rightarrow \mu^+\mu^- X$
had masked the leading order contribution
 from $W^+ s\rightarrow 
c$.[\citenum{NLO}]  The abundance of gluons in 
the nucleon enhances the importance of
$W^+ g\rightarrow c \bar s$.  The old analysis
 of the strange quark momentum fraction,
which gave a value around 2.6\%, omitted a negative
 NLO contribution from $W^+ g$ fusion
which masks an even larger strange quark 
fraction.  
\item It has long been hoped that the 
right model of ``constituent'' quarks,
superpositions of the ``current'' quarks 
that are measured in electroweak
processes, could account for the main features 
of OZI rule violation.  Constituent
quarks, $Q$, would be superpositions of current 
quarks, $\bar q q $ pairs, and
gluons.  Different dynamical pictures of the 
transformation from $q$ to $Q$
suggest different strangeness mixing patterns.
Consider the $m_s=0$ limit.  Perturbative
QCD would suggest,
\begin{equation}
	u\rightarrow u (\bar u u + \bar d d + \bar s s)^n \Rightarrow U,
\end{equation}
{\it etc.\/} because pairs created by gluons are flavor singlets.  A 
non-perturbative redefinition of the quark fields occurs at the scale of 
chiral symmetry breaking.  The flavor structure of constituent quarks can 
depend on the dynamics that drives that transition.  In the
Nambu-Jona Lasinio model of 
chiral symmetry breakdown through $\bar q q \bar q q$
operators each quark flavor mixes only with its own type,
\begin{equation}
	u\rightarrow u (\bar u u)^n \Rightarrow U,
\end{equation}
{\it etc.\/} because $SU(3)$ symmetric four quark operators only involve one
flavor.  On the other hand 
anomaly (or instanton) induced dynamical chiral symmetry
breakdown gives
\begin{equation}
	u\rightarrow u (\bar d d \bar s s)^n \Rightarrow U,
\end{equation}
because of the determinental character of the 't Hooft
interaction.[\citenum{BJM}] Can the study of nucleon strange quark
matrix elements establish the usefulness of the notion of a
current$\rightarrow$constituent transformation and help decide the
dynamical mechanism behind it?
\item  An impressive list of experiments 
are underway to probe strangeness matrix
elements not yet measured.  For a thorough discussion and references to the
original proposals see the review in Ref.~[\citenum{DonMus}].   
Here is a short list ---
\begin{itemize}
	\item LSND underway at Los 
	Alamos hopes to measure $\Delta s$ via the axial
	coupling of the $Z^0$ in quasielastic $\nu p\rightarrow\nu p$ and $\nu
	n\rightarrow\nu n$ from carbon nuclei.  
	\item SAMPLE underway at Bates expects 
	to get a sensitivity of $\pm 0.2$ in $\mu_s$
	by measuring $A_{LR}$ in $\vec e p\rightarrow e p$ 
	and $\vec e d\rightarrow e d$ at $E_e=200 MeV$.
	\item At least three experiments being developed at CEBAF:  In Hall-A,
	E91-010 plans a 5\% measurement of $A_{LR}$ in $\vec e
	p\rightarrow e p$ at the level of
	$2\times 10^{-5}$, seeking to measure 
	both $\mu_s$ and $\langle r_s^2\rangle$.  In
	Hall-C, E91-017 plans an approximately 5\% 
	measurement of the combination
	$G_E^{(s)} + G_E^n + 0.2 G_M^{(s)}$ at 
	forward angles using $\vec e p\rightarrow e
	p$ and $\vec e d\rightarrow e d$. At backward angles the same 
	experiment plans to
	separate $G_E^{(s)}$ and $G_M^{(s)}$ at larger momentum 	
	transfer.  Also in Hall-A,
	E91-004 plans to study $A_{LR}$ in 	
	$\vec e\phantom{1}^4\!He\rightarrow e\phantom{1}^4\!He$ at $Q^2\approx
	0.6 GeV^2$.  Spinless $\phantom{1}^4\!He$ removes magnetic and axial 
	vector effects.  The
	experiment should be sensitive to $\langle r^2_s\rangle$.  
	With $\langle
	r^2_s\rangle = 0$, 
	$A_{LR}\approx 5\times 10^{-5}$.  The experiment aims for a
	sensitivity of 30\% of this value.
	\item Proposal A4/1-93 for the Mainz 
	electron accelerator aims to measure
	$\langle r^2_s\rangle$ via $A_{LR}$ in $\vec e 
	p\rightarrow ep$ at $E_e=855 MeV$.
	\item Finally, both the SMC 	
	group at CERN and Hermes at HERA plan to try to
	separate quark flavor contributions $g_1$ making use of fragmentation
	triggers.[\citenum{Mank}]  A draft of early SMC results has already appeared.
	[\citenum{Wis}]
\end{itemize}
With time a more complete picture of the 
strangeness contamination of the nucleon
seems destined to emerge from this 
wide range of efforts at many laboratories.
\end{itemize}

\subsection{Transverse Spin}

Today's excitement about polarized hadronic collider physics is intimately
connected with the revival of interest in transverse polarization effects in deep
inelastic processes.  In the early days of QCD when only deep inelastic lepton
scattering (DIS) had been studied in detail, longitudinal and transverse
polarization effects were thought to be quite different.
[\citenum{Feyn}]  Longitudinal
asymmetries in DIS are described by the twist-two (scaling) structure function
$g_1$, while transverse asymmetries measure the twist-three (${\cal O}(1/\sqrt{Q^2})$)
structure function $g_2$.  Twist-three structure functions are interaction
dependent, and do not have simple parton interpretations.  This led to the
erroneous impression that transverse 
spin effects were inextricably associated with off-shellness, transverse
momentum and/or quark-gluon interactions.  Now we understand that there is a
twist-two quark distribution, $h_1(x,Q^2)$, describing transverse spin
effects,[\citenum{RS}] that it decouples from DIS due to a chirality selection rule of
perturbative QCD, and that it dominates transverse asymmetries in certain hard
processes such as Drell-Yan production of lepton
pairs.[\citenum{Artru,JafJi,RalCorPir}]  The correspondence between longitudinal and
transverse spin effects is completed by a twist-three, longitudinal structure
function, $h_2(x,Q^2)$, analogous to $g_2$.  The correspondence is summarized in
Table~\ref{tbl:trans}.
\begin{table}[ht]
	\center{
	\begin{tabular}{|c|c|c|}
	\hline
			 & Longitudinal & Transverse \\   
			 & Polarization & Polarization \\ \hline
	Twist-2          & $g_1(x,Q^2)$ & $h_1(x,Q^2)$\\ \hline
	Twist-3          & $h_2(x,Q^2)$ & $g_2(x,Q^2)$\\
	\hline
	\end{tabular} 
	       }
	\caption{{\sf The transverse and longitudinal
distribution functions through twist-three.}}
	\label{tbl:trans}
\end{table}

Experiments to measure $h_1$ include Drell-Yan 
with transversely polarized target
and beam --- $\vec p_\perp \vec p_\perp\rightarrow \bar\ell\ell+X$,
[\citenum{RS,JafJi}]
direct photon and jet production at large $p_\perp$ --- $\vec p_\perp \vec
p_\perp\rightarrow\pmatrix{jj\cr j\gamma} +X$,[\citenum{Jijet,JafSai}] 
and inclusive pion or
lambda production in deep inelastic scattering --- 
$\vec e \vec p_\perp\rightarrow
\pmatrix {\Lambda\cr\pi} +X$.[\citenum{Artru,JafJi2}]

As yet relatively little is known 
about the properties of $h_1$.  Here is a brief
summary.  More detail can be found in Ref.~[\citenum{Erice}]
\begin{itemize}
	\item $h_1$ measures the probability to find transversely 
	polarized quarks (along,
	say $\hat x$) in a transversely polarized 
	nucleon moving in the $\hat z$ direction
	--- known, for brevity, as ``transversity''.
	\item For non-relativistic dynamics $h_1=g_1$, so 
	the difference is a measure of
	relativistic effects.
	\item $h_1$, $g_1$ and $f_1$ obey inequalities, $\vert 
	h_1(x,Q^2) \vert\le
	f_1(x,Q^2)$, $\vert g_1(x,Q^2) \vert\le f_1(x,Q^2)$, and
	$f_1(x,Q^2)+g_1(x,Q^2)\ge 2\vert h_1(x,Q^2)\vert$ for 
	each flavor of quark and
	antiquark.  The last was recently 
	suggested by Soffer.[\citenum{Sof}]  It is only
	approximately true, since it suffers radiative 
	corrections.[\citenum{GJJ,Cont}]
	\item Model calculations suggest that 
	$h_1$ is of the same order as
	$g_1$.[\citenum{JafJi2,h1size}]
	\item $h_1$ is related to 
	``tensor operators'' in the same sense as $g_1$ is
	related to axial vector operators.  In 
	particular, the lowest moment of $h_1$
	obeys a simple sum rule,[\citenum{JafJi}]
	\begin{eqnarray}
	2S^j\delta q^a(Q^2) &\equiv &\langle 
	P,S\vert \bar q^a i\sigma^{0j}\gamma_5
	q^a\vert_{Q^2}\vert P,S\rangle {\rm and}\nonumber\\
	\delta q^a(Q^2) & = & \int_0^1 
	dx (h^a_1(x,Q^2) - \bar h^a_1(x,Q^2))\nonumber\\
	\end{eqnarray}
	for each quark flavor, $a$.
	\item  There is no gluon analog for $h_1$.  
	This has interesting consequences for
	ratios of transverse to longitudinal asymmetries 
	in polarized hadronic processes. 
	We save that discussion for the \S 4.
\end{itemize}

\subsection {Higher Twist}

Higher twist is the generic name for ${\cal O}(1/{Q^n})$ corrections to
deep inelastic processes.  If corrections to DIS and other hard
processes could be measured precisely enough
as functions of $x$ and $Q^2$ it would be
possible to map out interesting quark-quark 
and quark-gluon correlations within the
nucleon.  Although interest in higher twist 
effects has persisted since the mid
'70's, theoretical and experimental difficulties 
have frustrated efforts to measure
and interpret them.  A major difficulty 
is that in spin-average deep inelastic
scattering or annihilation corrections 
start at ${\cal O}(1/Q^2)$ and these
effects are always buried beneath 
dominant, scaling contributions.[\citenum{JS,EFP}] 
In addition, target mass corrections 
$\propto M^2/Q^2$ also have to be removed to
expose higher twist.  In order to isolate 
higher twist contributions to
$F_2$ for example, it is necessary 
first to fit the leading twist contribution which
varies like $(\ln Q^2)^\gamma$ in leading 
order and subtract it away.  Next target
mass corrections must be fit and subtracted 
away.  Small inaccuracies in the
subtraction procedure can mimic higher twist corrections.

One searches for higher twist at low-$Q^2$ 
where the corrections are large.  However
perturbative corrections to leading twist 
also vary rapidly at low-$Q^2$ and the
two cannot easily be distinguished.  Mueller, 
in particular, has emphasized the
ambiguity between higher twist and higher order 
radiative corrections to leading
twist.[\citenum{Mue}]  Consider, for example, a 
twist-four contribution ($\propto
\mu^2/Q^2$) in comparison with a 
perturbative correction to leading twist $\propto
(\alpha/\pi)^n$.  The two are 
indistinguishable over a small $Q^2$ interval when
$\left[{d\over {d\ln Q^2}}{\mu^2\over Q^2}\right]/{\mu^2\over Q^2} \approx 
{d\over {d\ln Q^2}}({\alpha\over\pi})^n/{\alpha\over\pi}$, 
or
${9n\alpha\over 4\pi} \approx 1$
The question of how to distinguish higher twist 
from higher order corrections to
leading twist is still unresolved.

Spin dependent effects provide an apparently unique reprieve from this
complexity.  For certain spin dependent observables, it is possible {\it 
kinematically\/} to eliminate leading twist 
so that a higher twist structure (or
fragmentation) function dominates and does not have to be extracted as a
correction.  The classic example is $g_2$.  If the polarized target in DIS is
aligned at exactly $90^\circ$ to the lepton beam then the cross section is
suppressed by $1/\sqrt{Q^2}$ relative to any other polarization direction, and the
leading behavior measures the twist three structure function $g_2$.  $g_2$ and
$h_2$ seem to be unique in this regard.  $g_2$ dominates $\vec e \,\vec
p_\perp\rightarrow e + X$, while $h_2$ contributes dominantly to $\vec
p_\parallel\,\vec p_\perp\rightarrow\ell^+\ell^- +X$.

The twist-three, transverse spin structure function, $g_2$, deserves special 
attention because the first experimental data on $g_2$ have just been 
published.[\citenum{E143g2,SMCg2}]  A more extensive review of the properties and 
puzzles of $g_2$ can be found in Ref.[\citenum{RLJg2}].  Here is a brief summary 
of its most interesting features.
\begin{itemize}
\item $g_2$ consists of two pieces, one is a kinematic reflection of $g_1$ 
first noted by Wandzura and Wilczek.[\citenum{Wand}]
\begin{equation}
	g_2(x,Q^2) \equiv g_{2WW}(x,Q^2)+\bar g_2(x,Q^2),
\end{equation}
where
\begin{equation}
	g_{2WW}(x,Q^2) = -g_1(x,Q^2)+ \int_x^1 {dy\over y} g_1(y,Q^2)
\end{equation}
and $\bar g_2$ is the true twist three part of $g_2$.
\item $\bar g_2$ measures quark-gluon correlations in the nucleon.  It's
moments  are related to specific local quark-gluon operator products, such as
\begin{equation}
        \int _0^1 x^2 \bar{g}_2(x,Q^2) \propto
        \langle P,S \vert {1\over 8} g {\cal S}_{\mu_1 \mu_2} \bar{\psi}
        \tilde{G}_{\sigma \mu_1} \gamma _{\mu_2} \psi \vert P,S\rangle.
\label{eq:g2m2}
\end{equation}
Model builders or lattice enthusiasts who want to predict $g_2$ must confront 
such matrix elements.  
In other formulations of higher twist physics,
$\bar g_2$ can appear to depend on quark 
transverse momentum.[\citenum{RR}]  These different expressions are equivalent to 
eq.~(\ref{eq:g2m2}) by use of the QCD equations of motion, although the 
form may lead model builders in different directions.

\item  $g_2$ obeys an interesting sum rule first derived by Burkhardt and 
Cottingham,[\citenum{BurkCot}]
\begin{equation}
	\int_0^1 dx g_2(x,Q^2) = 0
	\label{eq:bcsr}
\end{equation}
The sum rule is a consequence of rotation 
invariance and an assumption about the good high energy behavior of Compton 
amplitudes.  It is easily derived from consideration of the bilocal operator 
matrix element that defines $g_1$ and $g_T\equiv g_1+g_2$,[\citenum{Erice,JafJig2}]
\begin{equation}
	\int {d\lambda\over{2\pi}}e^{i\lambda x}
        \left<PS\vert\bar\psi(0)\gamma_\mu\gamma_5\psi
        (\lambda n)\vert PS\right>  =   
        2\left\{ g_1(x) p_{\mu} S\cdot n + g_T(x)S_{\perp\mu}
        + M^2 g_3(x) n_\mu S\cdot n \right\}.
\label{eq:axial}
\end{equation}
If we work in the rest frame, integrate over $x$ and
take $\vec S \Vert \hat e_3$ and $\mu=3$, we find
        $\int\limits_{-1}^1 dx g_1(x,Q^2)=\langle P \hat e_3\vert\bar
        q(0)\gamma^3\gamma_5 q(0)\vert_{Q^2}\vert P \hat e_3\rangle.$
Next repeat the process with $\vec  S\parallel \hat e_1$ and $\mu=1$, with the
result,                                                                        
       $ \int\limits_{-1}^1 dx g_T(x,Q^2)=\langle P \hat e_1\vert\bar
        q(0)\gamma^1\gamma_5 q(0)\vert_{Q^2}\vert P \hat e_1\rangle.$
The right hand sides of these two equations are equal in the rest frame
by rotation invariance, whence the sum rule, 
apparently a consequence of rotation invariance.
The subtlety in this derivation is that the integral goes
from $-1$ to $1$ including $x=0$.  
$g_2(x,Q^2)$ is the limit of a function of $Q^2$ and $\nu$ and
therefore might contain a distribution ($\delta$--functions, {\it etc.\/}) at
$x=0$.  Since experimenters cannot reach $x=0$, the BC sum rule reads
\begin{equation}
        \int_0^1 dx g_2^{\rm observable}(x,Q^2) = -{1\over 2}c,
\end{equation}
where $c$ is the coefficient of the $\delta$-function.
This pathology is not as arbitrary as it
looks.  Instead it is an example of a disease known as a {\it``$J=0$ fixed 
pole with non-polynomial residue''.\/} First studied in Regge
theory,[\citenum{FoxFr70,BGJ}] a $\delta (x)$
in $g_2(x,Q^2)$ corresponds to a {\it real constant term in a spin flip
Compton amplitude which persists to high energy\/}.
There is no fundamental reason to exclude such a constant.
On the other hand the sum rule is known to be satisfied in QCD perturbation
theory through order $O(g^2)$. The sum rule has been studied by       
several groups who find no evidence for a
$\delta (x)$ in perturbative QCD.[\citenum{pertg2}]
So at least provisionally, we must regard this as a reliable sum rule.
At least one other sum rule of interest experimentally, the Gerasimov, Drell,
Hearn Sum Rule for spin dependent Compton scattering has the same potential
pathology. For further discussion of the BC sum rule see Ref.~[\citenum{RLJg2}]

\item It has taken a long time to figure out the evolution of twist three 
structure functions such as $g_2$.  The subject is too complex and too 
theoretical to be covered here.  The interested reader can find a review in 
Ref.~[\citenum{Erice}] and in recent talks by Kodaira and Uematsu.[\citenum{g2evol}] 
Roughly speaking, knowing $g_2$ as a function of $x$ at a  given $Q^2$ is not
sufficient to determine it at larger $Q^2$.   There are a couple 
of distribution functions of two momentum fractions, $x$ and $y$, $G(x,y,Q^2)$ 
and $\tilde G(x,y,Q^2)$, which evolve in the standard fashion with GLAP 
splitting functions, {\it etc.\/}  $G$ and $\tilde G$ probe correlated 
quark-gluon distributions in an infinite momentum frame.  $g_2$ is obtained 
from $G$ and $\tilde G$ by integrating out the $y$ dependence, but this 
integrates out information necessary for evolution.  Several years ago Ali, 
Braun and Hiller[\citenum{ABH}] pointed out that these problems are less severe 
at large $N_c$ and large $x$, and more recently, 
Strattmann[\citenum{Stratt}] used their methods to evolve bag predictions for 
$g_2$[\citenum{JafJig2}] to large enough $Q^2$ to compare with data.
\item Model predictions for the lowest non-trivial moment of $g_2$, 
eq.~(\ref{eq:g2m2}) differ significantly in sign and 
magnitude.[\citenum{JafJig2,QCDSR}]  
\end{itemize}

\begin{figure}
        \centerline{\epsfxsize=3in\epsffile[108 335 462
719]{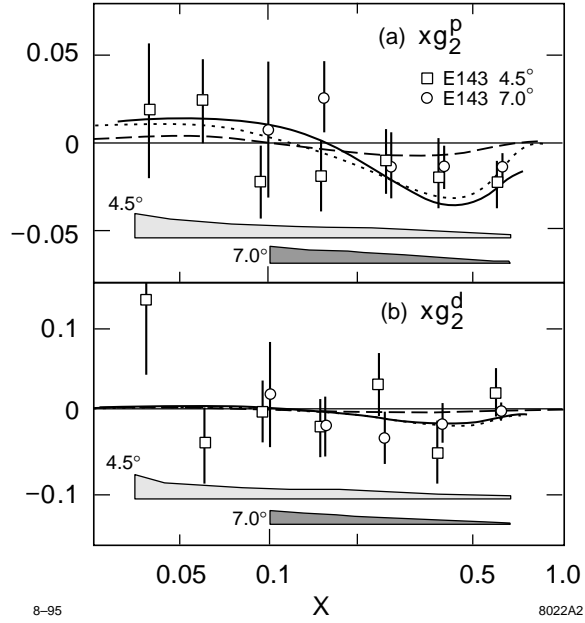}}
        \caption{{\sf SLAC E-143 data on $g_2$.}}
        \label{fig:E143g2}
\end{figure}

Early data on $g_2$ provided only upper limits.  The E143 Collaboration at 
SLAC recently published the first non-zero measurements, shown in 
Fig.~(\ref{fig:E143g2}).  This looks like the beginning of a significant
program in twist three physics.

\section{Physics Prospects at Polarized Hadron Colliders}

Nucleon spin and flavor physics has become a major focus of particle and 
nuclear physics programs.  The physics issues outlined in the previous section 
can be addressed at ongoing and planned experiments at CERN (HMC studying 
$\vec \mu\vec N \rightarrow  \mu h X$), SLAC (the E150's studying $\vec e\vec 
N\rightarrow e X$), HERA (Hermes studying $\vec e \vec N \rightarrow e X$,
$\vec e \vec N  \rightarrow  e h X $, and
$\vec e \vec A  \rightarrow  e X $)
and CEBAF (the many experiments studying $\vec e \vec p \rightarrow e p$).
Other more speculative proposals abound.  These include a dedicated European 
electron facility (ELFE), proposals to polarize the main injector at Fermilab 
and the proton ring at HERA.  There are so many possibilities to discuss --- 
they will dominate the program of this workshop --- that I will restrict 
myself to a few that are particularly 
well matched to a polarized hadron collider, and contrast them with 
experiments seeking the same information at some of the other facilities 
listed above.

The list of {\it desiderata\/}, particularly well suited to hadron colliders 
includes measurements of the longitudinal polarization asymmetries, 
$\Delta G$, $\Delta\bar u$, $\Delta\bar d$, $\Delta u$ and $\Delta d$, and the 
transversity distribution, $h_1$.  

There are good, specific reasons to turn to polarized colliders as a source of 
new information on hadron structure.  
\begin{itemize}
\item  First, chiral odd processes are not 
excluded, so quark transversity distributions can be measured.  
\item Second, a greater diversity of flavor probes is 
available.  Drell-Yan processes with 
weak bosons such as $\vec p \vec p \rightarrow Z^0 + X \rightarrow \ell
\bar\ell + X$ and $\vec p \vec p \rightarrow W^\pm + X \rightarrow \ell
\bar\ell + X$ complement $\vec p \vec p \rightarrow \gamma + X \rightarrow \ell
\bar\ell + X$ and provide flavor information that would require the 
unrealizable process $\nu \vec p 
\rightarrow \pmatrix{\mu\cr\nu} + X$ in the lepton-hadron scattering 
domain.  
\item Finally, it is easier to exploit the gluon-gluon and quark-gluon 
coupling to obtain information about polarized gluon distributions.  Gluons do 
not couple to leading order in deep inelastic lepton scattering.  They appear 
either through evolution or in higher order processes like heavy quark pair 
or two jet production.  In a polarized hadron collider polarized gluons should 
dominate the most copious hard processes like two jet production.
\end{itemize}

\subsection{Transverse Versus Longitudinal Asymmetries --- A Test of the QCD 
Parton Formalism}

Sometimes we are so busy looking for ways to measure small and interesting 
effects that we neglect the obvious:  
{\it Transverse asymmetries in deep inelastic 
hadron-hadron collisions (with the important exception of Drell-Yan processes)
are extraordinarily small because there is no gluon transversity distribution 
at leading twist,\/}  
\begin{equation}
{{\cal A}_{TT}\over {\cal A}_{LL}}\ll 1
\end{equation}
for processes such as $pp\rightarrow jj + X$, $pp\rightarrow \gamma j +X$, 
$pp\rightarrow \pi(k_\perp) + X$, {\it etc.\/} ($\pi(k_\perp)$ denotes a pion 
produced at large $k_\perp$).[\citenum{Jijet,JafSai}]

The argument for this suppression involves the following steps. 
\begin{itemize}
\item ${\cal A}_{TT}$ measures products 
of parton transversity distributions times fundamental parton-parton 
scattering cross sections.  
\item $gq\rightarrow gq$ and 
$gg\rightarrow gg/q\bar q$ dominate jet production in hadron-hadron collisions 
except at very large $z$.  
\item There is no gluon transversity distribution at twist-two.  The 
transversity distribution is the imaginary part of helicity flip 
parton-nucleon forward scattering.  Leading twist gluons have helicity $\pm 
1$, and cannot flip their helicity ($\Delta$ helicity $= \pm 2$)
scattering forward from a nucleon which can only absorb $\Delta$ helicity 
$=\pm 1$.  So gluons decouple from transverse asymmetries.  On the other hand,

\item There is every reason to expect a significant polarized gluon 
distribution in the nucleon ($\Delta G \approx G$), so longitudinal 
asymmetries in jet production should not be small.

\item Of the terms that might be large only $qq\rightarrow qq$ remains.  
However it has long been known that this contribution to ${\cal A}_{TT}$ is 
suppressed by a color exchange factor of $\sim {1\over 11}$.[\citenum{HMS,Jijet}]
\end{itemize}
The ratio ${{\cal A}_{TT}\over {\cal A}_{LL}}$ is 
likely to be so small that it will not be 
observed until polarized hadron colliders have run for many years.  Detailed 
predictions for specific processes will soon be available.[\citenum{JafSai}]

This prediction tests the whole parton spin formalism.  
In the days before the discovery of $h_1$, $g_2$ was believed to be the 
distribution function relevant to transverse asymmetries.  Since gluons {\it
do \/} couple to $g_2$ there would be no reason for ${\cal A}_{TT}$ to be
suppressed.   Indeed, the first paper to discuss suppression 
of ${\cal A}_{TT}$,[\citenum{HMS}] uses 
$g_2$, includes gluons, and drops them later, 
presumably under the assumption 
that gluons are not polarized. 

It is interesting to contrast polarized Drell-Yan processes such as $\vec p 
\vec p\rightarrow \ell\bar\ell + X$ which must proceed in leading order by 
$\bar q q$ annihilation to leading order.  
In this case gluons do not participate in either 
the transverse or longitudinal asymmetry.  
So $\left[{\cal A}_{TT}\over {\cal A}_{LL}
\right]_{DY}\approx 1$ unless the quark or antiquark transversity 
distributions are small 
compared to their helicity distributions.  

This selection rule can only be tested at a polarized hadron collider.  It is 
relatively independent of beam polarizations (which are hard to measure 
absolutely), and experimental acceptances.  It should be a high
priority.

\subsection{A Brief Summary of Some Flagship Experiments}

Much of this Workshop will be devoted to the introduction and study of a few 
flagship experiments at a high energy polarized collider.  I will only list 
them and their competition at other facilities, leaving the full presentations 
to later speakers.

\subsubsection {$\Delta G$}
The measurement of the polarized gluon distribution in the nucleon is perhaps 
the most important immediate goal in all of hadron spin physics.  Because 
hadrons contain a lot of glue, polarized hadron collisions are a natural place 
to look for the effects of $\Delta G$.  Two processes look promising:  $\vec p 
\vec p \rightarrow \gamma + {\rm jet}$ and $\vec p \vec p \rightarrow {\rm 2 
jets}$.  The contributing graphs in leading order are all variants of QCD 
Compton scattering and they measure either $\Delta G \otimes \Delta q$ or 
$\Delta G \otimes \Delta G$.  Good jet reconstruction and good electromagnetic 
calorimetry are priorities.

Competition from other regimes include careful evolution studies of 
$g_1^{eN}(x,Q^2)$ and electroproduction of $\bar cc$ pairs.  Gluons do not 
couple directly into deep inelastic scattering but they effect the evolution 
of the quark distribution via the GLAP equations.  Schematically,
\begin{equation}
{d\over d\ln Q^2}\: g_1^{eN} \propto {\alpha_s\over\pi}\{ {\cal 
P}_{QQ}\otimes g_1^{eN} \, + \, {\cal P}_{QG} \otimes \Delta G\}.
\label{eq:glap}
\end{equation}
Sufficiently accurate measurements of the $Q^2$ 
dependence of $g_1^{eN}$ should allow the 
extraction of $\Delta G(x,Q^2)$.  Indeed some workers believe they see 
evidence of a large positive $\Delta G$ in existing data.[\citenum{BFR}]  One 
difficulty of this method is its scheme dependence:  there are 
perfectly reasonable schemes (like $\overline{MS}$) where the integrated 
$\Delta G$ decouples from deep inelastic scattering and GLAP evolution.
The $\bar cc$ option, first proposed by Carlitz, Collins and 
Mueller,[\citenum{gluon}] is based 
on photon--gluon fusion and could be attempted
at  electron facilities.
\subsubsection{$\Delta\bar q$}
If one accepts the now standard analysis of CERN and SLAC data, then 
antiquarks are probably, but not certainly, polarized.  Consider, for example, 
strange quarks.  We know that the nucleon contains both $s$ and $\bar s$ 
quarks ({\it e.g.} from $\pmatrix{\nu\cr\bar\nu}p\rightarrow \mu^+\mu^- X$), 
and we know that $\Delta s+\Delta \bar s\ne 0$.  
 [For the moment we use 
$\Delta q$ and $\Delta \bar q$ to refer to the helicity asymmetry of quarks 
and antiquarks respectively.]  Still, this is not enough to
prove that $\Delta\bar s\ne 0$.  Neutrino
scattering from  polarized targets is out of the question, but the analog
Drell-Yan process is  a natural goal for a polarized hadron collider.  Consider
$\vec p p\rightarrow  W^\pm + X$, a parity violating process that proceeds
dominantly via $u \bar  d\rightarrow W^+$ or $d \bar u \rightarrow W^-$.  In
the parton model,
\begin{equation}
{\cal A}^+_L \sim {\Delta u(x_1) \bar d(x_2) - \Delta \bar d(x_1) u(x_2)\over
u(x_1)\bar d(x_2) + \bar d(x_1) u(x_2)}
\label{eq:asym}
\end{equation}
for $W^+$ and similarly for $W^-$. 
The interpretation is cleanest at large 
$x_2$ where ${\cal A}^+_L\Rightarrow \Delta\bar d(x_1)/\bar d(x_1)$.  A 
detector with good lepton, jet and missing transverse energy detection can 
make an accurate measurement of $\Delta \bar d/\bar d$ and $\Delta \bar u/\bar 
u$.

Competition comes mainly from the use of flavor triggers to identify the 
scattered quark in deep inelastic electron scattering, $\vec e \vec 
p\rightarrow e + \{\pi^\pm,\pi^0, K^\pm\} + X$.  Such studies will be 
undertaken by HMC (a successor to SMC at CERN) and by Hermes at HERA, but 
difficulties separating current and target fragments, and the unknown 
efficiency of flavor tagging via fragmentation will make this a difficult 
exercise.

\subsubsection{$\Delta q$}

Clearly, quark helicity distributions can be extracted at the other kinematic 
limit of eq.~(\ref{eq:asym}), when $x_1$ is large and $x_2$ is near zero.  
Then ${\cal A}^+_L\sim \Delta u(x_1)/u(x_1)$.  Once again competition appears 
to be limited to flavor tagging experiments at electron (or muon) facilities.

\subsubsection{Transversity}

Polarized hadron colliders offer the possibility of measuring the nucleon's 
third (and final) twist two quark-parton distribution, the transversity, 
$\delta q(x,Q^2)$.  It will not be 
easy.  Transversity effects in two jet production and $\gamma$ plus jet are 
very small.  Drell-Yan looks better but requires non-vanishing {\it 
anti\/}--quark transversity, $\delta\bar q\ne 0$.  Although the process is 
clean in theory, it makes heavy demands on detectors (lepton identification, 
large acceptance, $\ldots$) and accelerator (high luminosity, high 
polarization, good polarimetry)
and will not be the first measurement made at a high energy polarized 
collider.  The prospects for measuring transversities at other facilities are 
also challenging.  I know of three classes of proposals:  (1)~$\vec e \vec 
p_\perp \rightarrow e'\vec\Lambda_\perp X$, where the self-analyzing 
decay of the $\Lambda$ allows extraction of the product of the nucleon 
transversity multiplied by an analogous $\Lambda$--fragmentation 
function.[\citenum{Artru}]
A particular shortcoming of this proposal is that only strange quarks are 
likely to have a strong transversity correlation between the target nucleon 
and the $\Lambda$.  (2)~ $\vec e \vec p_\perp\rightarrow e'\pi\pi X$, 
where the pions are used to construct a jet variable correlating with the 
target transversity distribution.  Our lack of understanding of the analyzing 
power of the multipion jet fragmentation process complicates this method.  
(3)~Finally $\vec e \vec p_\perp\rightarrow e'\pi X$ has no leading twist 
asymmetry at all.[\citenum{JafJi}]  At twist three there are two contributions, one 
of which is proportional to the target transversity times a twist three, spin 
average pion fragmentation function.  A lot of unraveling would be required 
to extract transversity distributions from this process.  It could be 
undertaken at HMC and Hermes.

The prospects for each of these projects at a high energy polarized collider 
will be discussed in much more detail later in this Workshop.  

\section{Acknowledgments}
I would like to that X.~Ji, N.~Saito, and M.~Tannenbaum for discussions 
relating to this work.

\bibliography{ctp2518refs}
\end{document}